\documentclass{aastex631}
\usepackage{hyperref}

\begin{document}

\title{Photometric observations of flares on AD Leo from GWAC-F30 and TESS} 

\correspondingauthor{Jian-Ying Bai}
\email{baijianying@nao.cas.cn}

\author{Jian-Ying Bai}
\affiliation{Key Laboratory of Space Astronomy and Technology, National Astronomical Observatories, Chinese Academy of Sciences, Beijing 100101,	People's Republic of China}

\author{Jing Wang}
\affiliation{Key Laboratory of Space Astronomy and Technology, National Astronomical Observatories, Chinese Academy of Sciences, Beijing 100101,	People's Republic of China}
\affiliation{Guangxi Key Laboratory for Relativistic Astrophysics, School of Physical Science and Technology, Guangxi University, Nanning 530004, People’s Republic of China}

\author{Hua-Li Li}
\affiliation{Key Laboratory of Space Astronomy and Technology, National Astronomical Observatories, Chinese Academy of Sciences, Beijing 100101,	People's Republic of China}

\author{Li-Ping Xin}
\affiliation{Key Laboratory of Space Astronomy and Technology, National Astronomical Observatories, Chinese Academy of Sciences, Beijing 100101,	People's Republic of China}

\author{Guang-Wei Li}
\affiliation{Key Laboratory of Space Astronomy and Technology, National Astronomical Observatories, Chinese Academy of Sciences, Beijing 100101,	People's Republic of China}

\author{Yuan-Gui Yang}
\affiliation{School of Physics and Electronic Information, Huaibei Normal University, Huaibei 235000, People’s Republic of China}

\author{Jian-Yan Wei}
\affiliation{Key Laboratory of Space Astronomy and Technology, National Astronomical Observatories, Chinese Academy of Sciences, Beijing 100101,	People's Republic of China}
\affiliation{School of Astronomy and Space Science, University of Chinese Academy of Sciences, Beijing, People’s Republic of China}

\begin{abstract}
We observed active M dwarf star AD Leo for 146 hr in photometry by GWAC-F30 and also analyzed 528-hr photometric data of the star from TESS. A total of 9 and 70 flares are detected from GWAC-F30 and TESS, respectively. Flare durations, amplitudes and energies are calculated. The distributions of the three properties and FFDs are given. Within the same energy range of flares, the FFDs of AD Leo obtained in this research and the previous study are basically consistent, which suggests that the magnetic activity of this star has not significantly changed compared to that decades ago. Comparing with the average FFD of M-type stars, AD Leo's FFD is twice higher, indicating that its magnetic activity is more active than that of the average level of the  M-type. Based on TESS light curve, AD Leo's rotation period is calculated as 2.21$^{+0.01}_{-0.01}$ day , supporting the result given in previous research. During the decay phase of the most energetic flare from TESS, we identified QPPs and determined a 26.5-min oscillation period, which is currently the longest period for AD Leo, suggesting that long periodic physical process existed during flare of this star.
\end{abstract}




\section{Introduction} \label{sec:intro}
Stellar flares are violent explosion phenomena which lasts from a few seconds to hours. During the explosion, a large amount of energies can be released over a wide wavelength range, from radio to X-rays \citep{2002A&A...394..653S,2006ApJ...647.1349O,2007ApJS..173..673W,2010ApJ...721..785O,2012ApJ...748...58D}. These energies are believed to come from magnetic reconnection occurring in the stellar corona analogy to the Sun \citep{2000IrAJ...27..117G,2005Ap.....48..279T,2010ARA&A..48..241B,2019ApJ...871L..26F}. Flares are not only important for understanding star formation and evolution, but may also help understanding the habitability evolution of planets around active stars \citep[e.g.,][]{2008ASPC..384..303L,2014OLEB...44..239L}.

Several studies have suggested that stellar flares may be a possible threat for a planet to maintain its habitable atmosphere  \citep[e.g.,][]{2010AsBio..10..751S,2015AsBio..15..119L,2019AsBio..19...64T}. \cite{2019AsBio..19...64T} noted that ultra-violet (UV) fluxes and energetic particles from frequent flares can deplete ozone obviously, leading to a potentially uninhabitable planet. The photoevaporation driven by extreme ultra-violet (EUV) and X-ray could lead to erosion of atmosphere on low-mass planets \citep{2013MNRAS.430.1247L,2013AsBio..13.1030K,2016MNRAS.460.1300E,2017ApJ...851...77R}. \cite{2018ApJ...867...71L} indicates that stellar flares could dominate the far ultra-violet (FUV) energy budget on M stars. In addition to flares, coronal mass ejections (CMEs) may also erode a planet's atmosphere, potentially lead to losing entire atmosphere of one planet orbiting M star \citep{2007AsBio...7..185L,2009A&ARv..17..181L,2014MNRAS.439.3225L,2021ApJ...916...92W}.

AD Leo (GJ 388, BD +20 2465) is a bright active flare star (V = 9.52 mag), with mass of 0.43 M$_{\sun}$, spectral type of M3 and distance of 4.9 pc \citep{1989ApJS...71..245K,2012PASP..124..545H,2013A&A...549A.109B,2014ApJ...781L...9B,2014AJ....148...91L}. Its radius is 0.436 $^{+0.049}_{-0.049}$ R$_{\sun}$ and effective temperature is 3414 $^{+100}_{-100}$ K \citep{,2016ApJ...822...97H}. The metal abundance is 0.28 $^{+0.17}_{-0.17}$ and rotation period is 2.23$^{+0.36}_{-0.27}$ days \citep{2012ApJ...748...93R,2012PASP..124..545H}. \cite{2008MNRAS.390..567M} suggests that AD Leo is seen nearly pole-on by a spectropolarimetric study.

In this work, AD Leo are monitored in photometry for 146 hours. The Transiting Exoplanet Survey Satellite \citep[TESS;][]{2014SPIE.9143E..20R} also observed this star from 2021-Nov-06 to 2022-Feb-26 photometrically. We analyzed these photometric data. Flares are identified, and their properties are analyzed and discussed. The paper is organized as follows. In Section 2, the observations and data reduction are outlined. In Section 3, we describe the analysis method of the data. The results and discussions are given in Section 4. In Section 5, we provide a summary.

\begin{table}[t]
	\begin{center}

		\begin{tabular}{cccccc}
			\hline\hline\noalign{\smallskip}
			Date &  Filter      & Start & End & Duration  \\
			(UT) &  (Johnson)    & (UT) & (UT) & (Hour)  \\
			\hline\noalign{\smallskip}
			2021 Dec 25	&	B	&	15:11:12	&	19:25:47	&	4.24	\\
			2021 Dec 27	&	B	&	15:52:33	&	22:04:51	&	6.2	\\
			2021 Dec 28	&	B	&	15:48:20	&	21:35:31	&	5.79	\\
			2021 Dec 29	&	B	&	16:16:00	&	21:43:41	&	5.46	\\
			2021 Dec 30	&	B	&	15:45:49	&	21:40:05	&	5.9	\\
			2022 Jan 01	&	B	&	16:54:24	&	20:19:06	&	3.41	\\
			2022 Jan 03	&	B	&	16:45:51	&	20:07:48	&	3.37	\\
			2022 Jan 05	&	B	&	12:55:30	&	21:19:12	&	8.39	\\
			2022 Jan 06	&	B	&	16:30:25	&	21:59:59	&	5.49	\\
			2022 Jan 07	&	B	&	17:34:25	&	21:59:10	&	4.41	\\
			2022 Jan 08	&	B	&	18:15:33	&	19:44:09	&	1.48	\\
			2022 Jan 10	&	B	&	15:31:14	&	20:04:12	&	4.55	\\
			2022 Jan 12	&	B	&	17:02:29	&	20:26:01	&	3.39	\\
			2022 Jan 15	&	B	&	15:33:53	&	21:01:31	&	5.46	\\
			2022 Jan 16	&	B	&	18:27:46	&	22:12:23	&	3.74	\\
			2022 Jan 17	&	B	&	17:05:07	&	21:30:55	&	4.43	\\
			2022 Jan 25	&	B	&	15:45:34	&	19:26:04	&	3.67	\\
			2022 Jan 26	&	B	&	16:30:47	&	21:37:48	&	5.12	\\
			2022 Jan 27	&	B	&	15:32:55	&	20:39:56	&	5.12	\\
			2022 Jan 28	&	B	&	16:26:53	&	21:33:55	&	5.12	\\
			2022 Jan 29	&	B	&	18:00:56	&	22:18:38	&	4.29	\\
			2022 Feb 01	&	B	&	17:33:00	&	22:01:00	&	4.47	\\
			2022 Feb 02	&	B	&	18:27:27	&	21:48:57	&	3.36	\\
			2022 Feb 03	&	B	&	13:20:59	&	20:11:12	&	6.84	\\
			2022 Feb 04	&	B	&	16:28:16	&	21:56:01	&	5.46	\\
			2022 Feb 07	&	B	&	14:12:53	&	21:04:26	&	6.86	\\
			2022 Feb 08	&	B	&	13:58:49	&	20:48:52	&	6.83	\\
			2022 Feb 09	&	B	&	14:50:22	&	21:36:20	&	6.77	\\
			2022 Feb 10	&	B	&	13:57:49	&	20:47:52	&	6.83	\\
			
			\noalign{\smallskip}\hline
		\end{tabular}
		\caption{ {Log of photometry by GWAC-F30} }\label{Tab:t1}
	\end{center}
\end{table}

\section{Observations and data reduction} \label{sec:Observation}

AD Leo (RA 10:19:36.28, Dec +19:52:12.01) was observed by a 30-cm telescope which is part of the Ground-Based Wide-Angle Camera (GWAC) network \citep{2021PASP..133f5001H}. GWAC network (GWAC-N) is part of the ground segment of the space-based multi-band astronomical variable objects monitor (SVOM) mission \citep{2016arXiv161006892W}. The 30-cm telescope (GWAC-F30) is located at the Xinglong Observatory of National Astronomical Observatories, Chinese Academy of Sciences (NAOC) \citep{2022RAA....22f5002L}. It has a focal ratio of F/3.6, and is equipped with a FLI camera (ProLine 16803) which contains a CCD with 4096 $\times$ 4096 pixels, providing a field of view of 1.9 $\times$ 1.9 degree$^2$. The pixel scale is 1.7 arcsec. A set of Johnson-Cousins UBVRI filters is equipped for broadband photometry. The limiting magnitude in B band can reach 13.5 mag with signal-to-noise ratio (SNR) = 20 for 120 s of exposure.

In this work, AD Leo was monitored with B filter for 29 nights, from December, 2021 to February, 2022, $\sim$146 hr in total. The photometry log is shown in Table 1, containing the observation date, filter applied, the start and end times of each observing session and the session duration. The exposure time is 120 s and the dead time between two adjacent exposures is about 4 s, leading to a cadence of 124 s for each observation session. 

The images of the photometric observations are reduced using standard IRAF \footnote{IRAF is distributed by the National Optical Astronomy Observatory, which is operated by the Association of Universities for Research in Astronomy, Inc., under cooperative agreement with the National Science Foundation.} routines, including bias subtraction, dark subtraction and flat correction. We choose 2MASS J10194483+1953309 (RA 10:19:44.84, Dec +19:53:30.81) and BD+20 2464 (RA 10:19:36.34, Dec +19:53:42.00) as comparison and check stars respectively to perform differential photometry. The top panel of Figure 1 shows all the light curves of AD Leo observed by GWAC-F30.

AD Leo was also observed by TESS \citep{2014SPIE.9143E..20R} with 2-min cadence in optical band during sector 45, 46 and 48. The TESS data is publicly available from the Mikulski Archive for Space Telescopes (MAST \footnote{https://archive.stsci.edu/missions-and-data/tess}). Because of big noise in sector 45 and 46, we only analysed the data of AD Leo from sector 48. The Simple Aperture Photometry (SAP) and PreSearch Data Conditioned (PDC$\_$SAP) data are provided in TESS products. The PDC$\_$SAP light curves are applied in this work, in which long-term trends are removed but keeping short period astrophysical signals. The bottom panel in Figure 1 presents TESS light curve in sector 48 for AD Leo.

\begin{figure}[ht]
	\centering
	\includegraphics[angle=0,width=150mm]{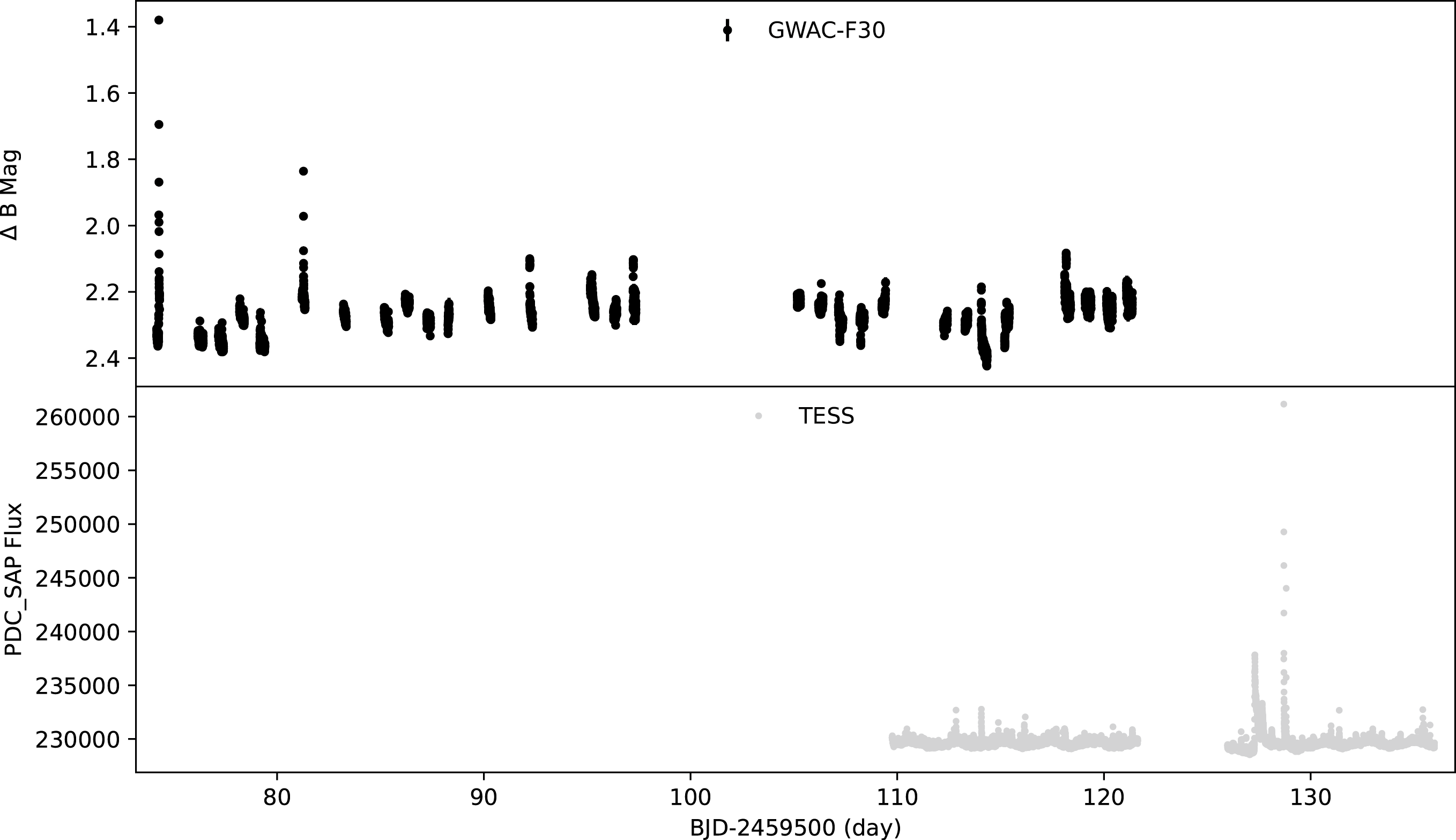}
	\caption{Light curves of AD Leo observed by GWAC-F30 and TESS. The top panel shows the data of GWAC-F30 in black points and the bottom presents that of TESS during sector 48 in gray. For both panels, the horizontal axis show the time in BJD - 2459500. The vertical axis of the top panel indicates differential magnitude in B band and that of the bottom is PDC$\_$SAP flux of TESS. \label{fig:f1}}
\end{figure}

\begin{figure}[ht]
	\centering
	\includegraphics[angle=0,width=150mm]{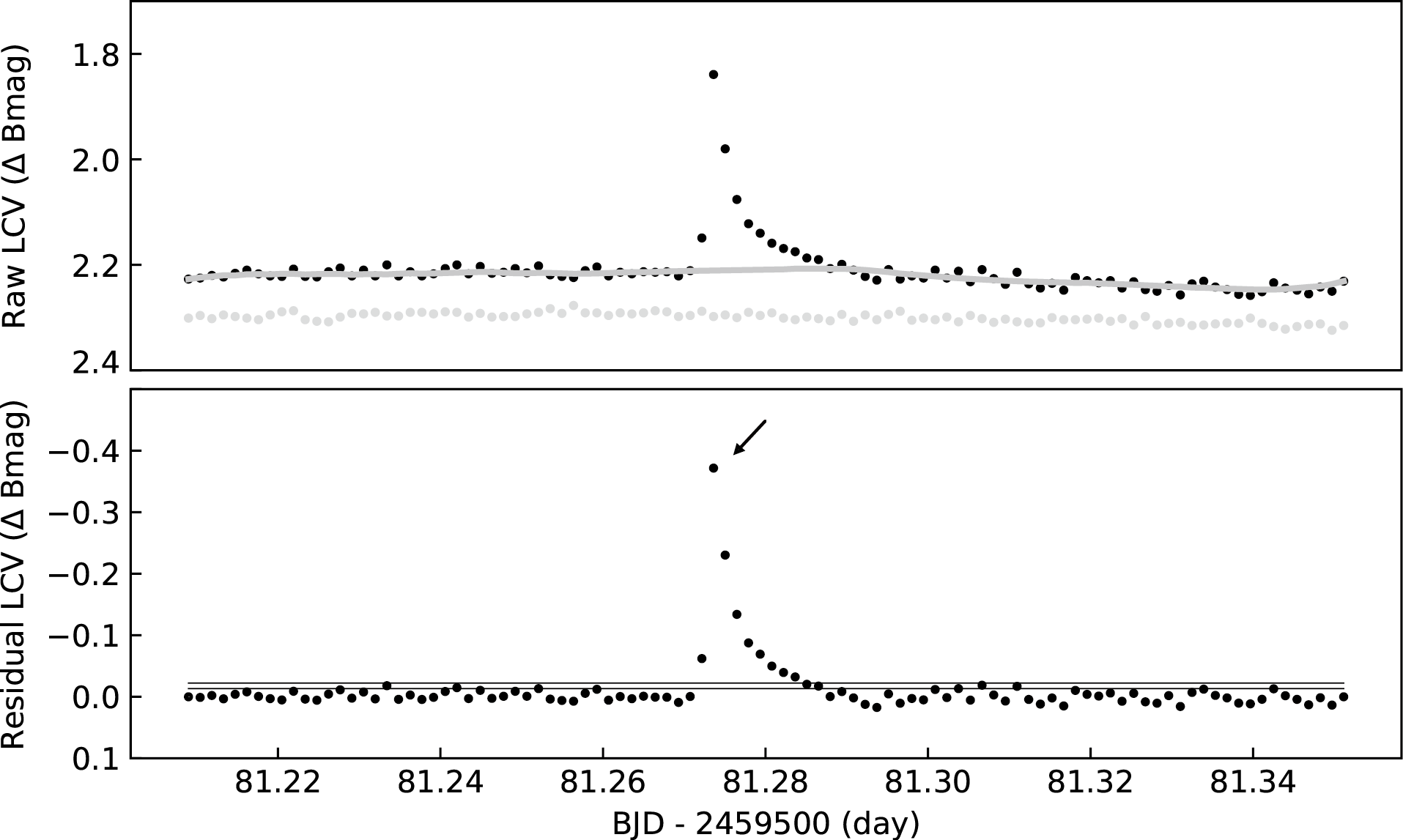}
	\caption{The Raw LCV and the Residual LCV on 2022 January 1. The light curves of AD Leo are marked with black points and the light curve of the reference star with gray. In the top panel, the thick gray line traces the fit of smoothing average for the quiescent phase. In the bottom panel, the two black lines indicate the positions of 3 $\times$ SD (lower) and 5 $\times$ SD (higher), respectively. The arrow points out the identified flare. \label{fig:f1}}
\end{figure}

\begin{figure}[ht]
	\centering
	\includegraphics[angle=0,width=150mm]{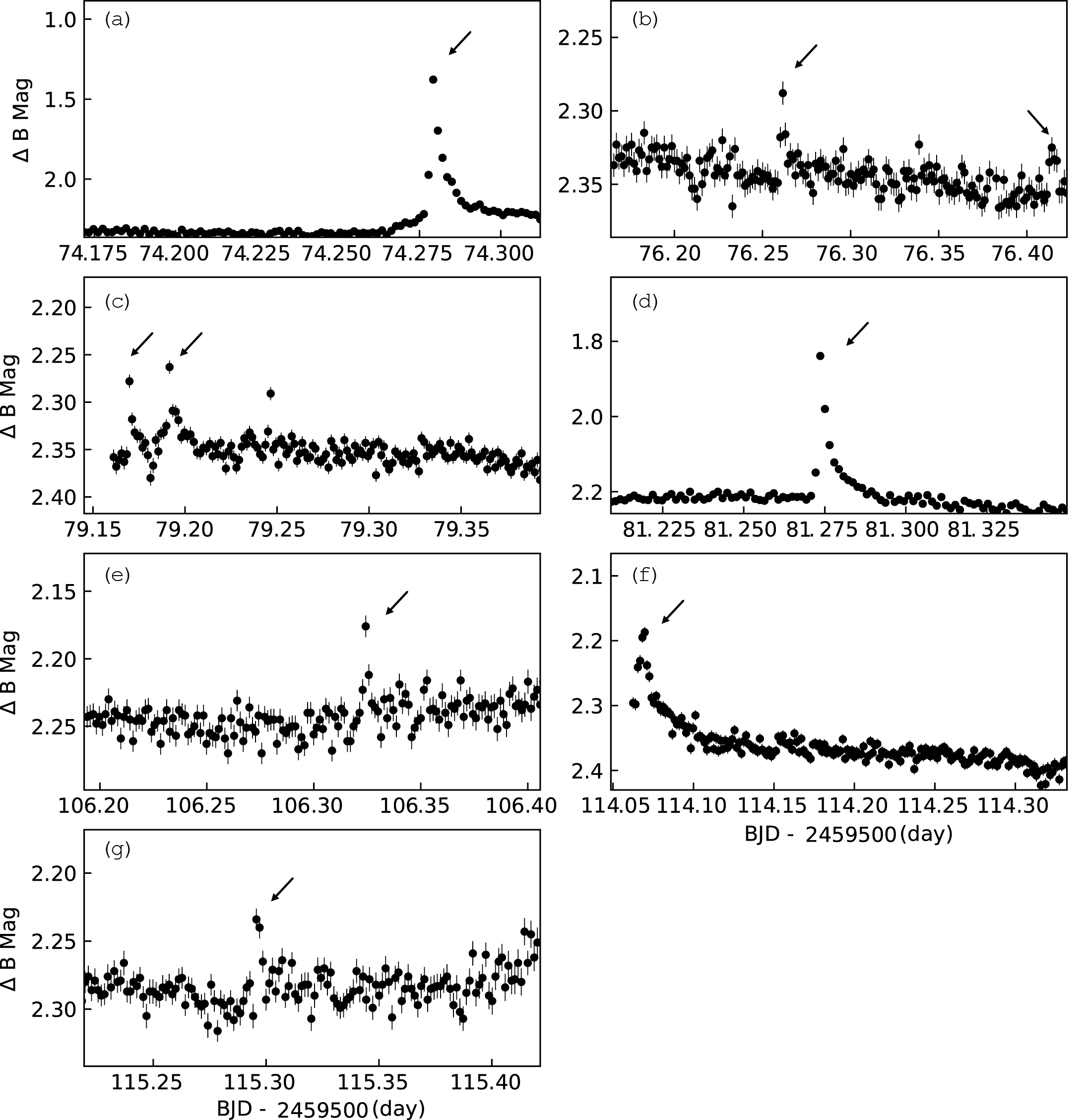}
	\caption{Light curves of the flares observed by GWAC-F30. The arrows indicate the flares. The error bars present the photometry errors and they are smaller than black points in panel a and d. \label{fig:f1}}
\end{figure}

\begin{figure}[h]
	\centering
	\includegraphics[angle=0,width=120mm]{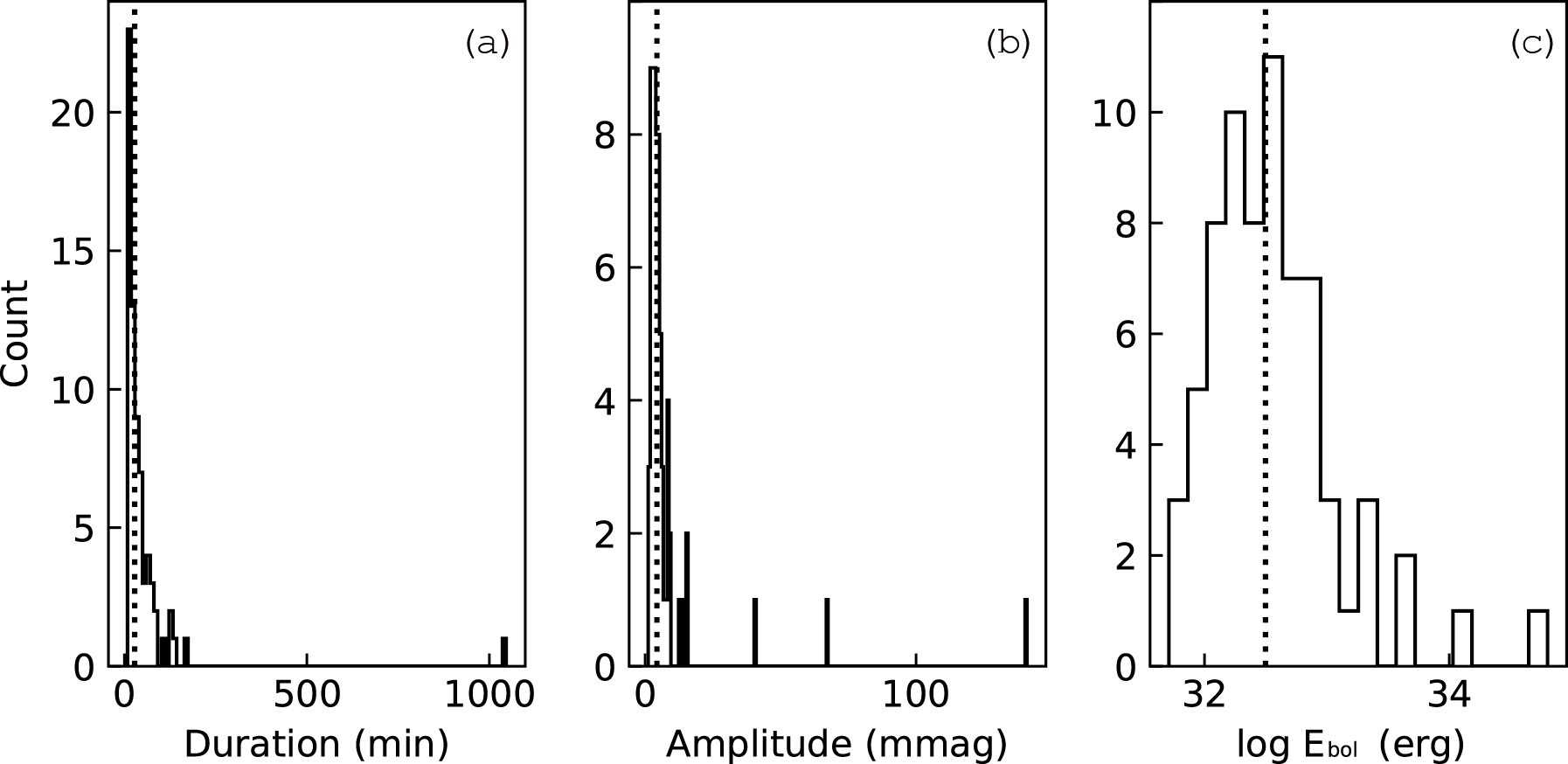}
	\caption{Histograms of flare durations, amplitudes and energies. The vertical dotted lines signify the medians of these parameters.\label{fig:f5}}
\end{figure}

\begin{table}[h]
	\begin{center}

		\begin{tabular}{cccccc}
			\hline\hline\noalign{\smallskip}
			Flare &  T$_{rise}$      & T$_{decay}$ & T$_{total}$ & Amplitude & log E$_{bol}$ \\
			ID &  (Minute)    & (Minute) & (Minute) & (mag) &  \\
			\hline\noalign{\smallskip}
			1	&	16.48	&	51.52	&	68.00	&	-0.960 	$\pm$	0.012	&	34.336	$\pm$	0.005	\\
			2	&	2.06	&	4.12	&	6.18	&	-0.055 	$\pm$	0.008	&	32.478	$\pm$	0.003	\\
			3	&	2.06	&	6.18	&	8.24	&	-0.031 	$\pm$	0.008	&	32.430	$\pm$	0.003	\\
			4	&	2.06	&	6.12	&	8.18	&	-0.074 	$\pm$	0.008	&	32.549	$\pm$	0.003	\\
			5	&	2.48	&	9.04	&	11.52	&	-0.082 	$\pm$	0.008	&	32.742	$\pm$	0.003	\\
			6	&	2.07	&	20.67	&	22.74	&	-0.372 	$\pm$	0.008	&	33.529	$\pm$	0.003	\\
			7	&	2.06	&	4.12	&	6.18	&	-0.063 	$\pm$	0.009	&	32.468	$\pm$	0.004	\\
			8	&	12.32	&	22.67	&	34.99	&	-0.17 	$\pm$	0.009	&	33.616	$\pm$	0.004	\\
			9	&	2.06	&	6.12	&	8.18	&	-0.054 	$\pm$	0.011	&	32.545	$\pm$	0.004	\\
			
			\noalign{\smallskip}\hline
		\end{tabular}
		\caption{ \centering {Properties of the flares detected by GWAC-F30} }\label{Tab:t33}
	\end{center}
\end{table}

\section{Data Analysis}\label{sec:Analysis}
The light curves of AD Leo from GWAC-F30 and TESS are analyzed to detect flares with a same program that has been applied in \cite{2021RAA....21....7B}. The processing method of the program is briefly described here, more details can be found in \cite{2021RAA....21....7B}. First, the light curve of AD Leo from single night (Raw LCV) is fitted iteratively to remove outliers by smoothing average, obtaining a quiescent phase \citep{2014ApJ...797..121H,2016ApJ...829L..31D,2017ApJ...849...36Y,2019ApJS..241...29Y}. Next, the Raw LCV is subtracted by the quiescent phase in order to get a residual light curve (Residual LCV). Finally, the Residual LCV is analyzed to identify flare candidates with the similar criteria used in previous studies (see Fig. 2) \citep[e.g.,][]{,2014ApJ...797..121H,2021RAA....21....7B}. The criteria are listed as follows: (1) The light curve of a flare candidate contains three consecutive measurements; (2) The measurements are more than three times the standard deviation (SD) of the quiescent phase, and at least one of them is more than five times the SD. After applying the criteria, the profile of the light curve for each candidate is also examined by visual inspection to confirm it consists of an impulsive rise (relatively short) and an exponential decay (relatively long). Figure 3 presents the flare light curves observed by GWAC-F30.

\begin{figure}[h]
	\centering
	\includegraphics[angle=0,width=110mm]{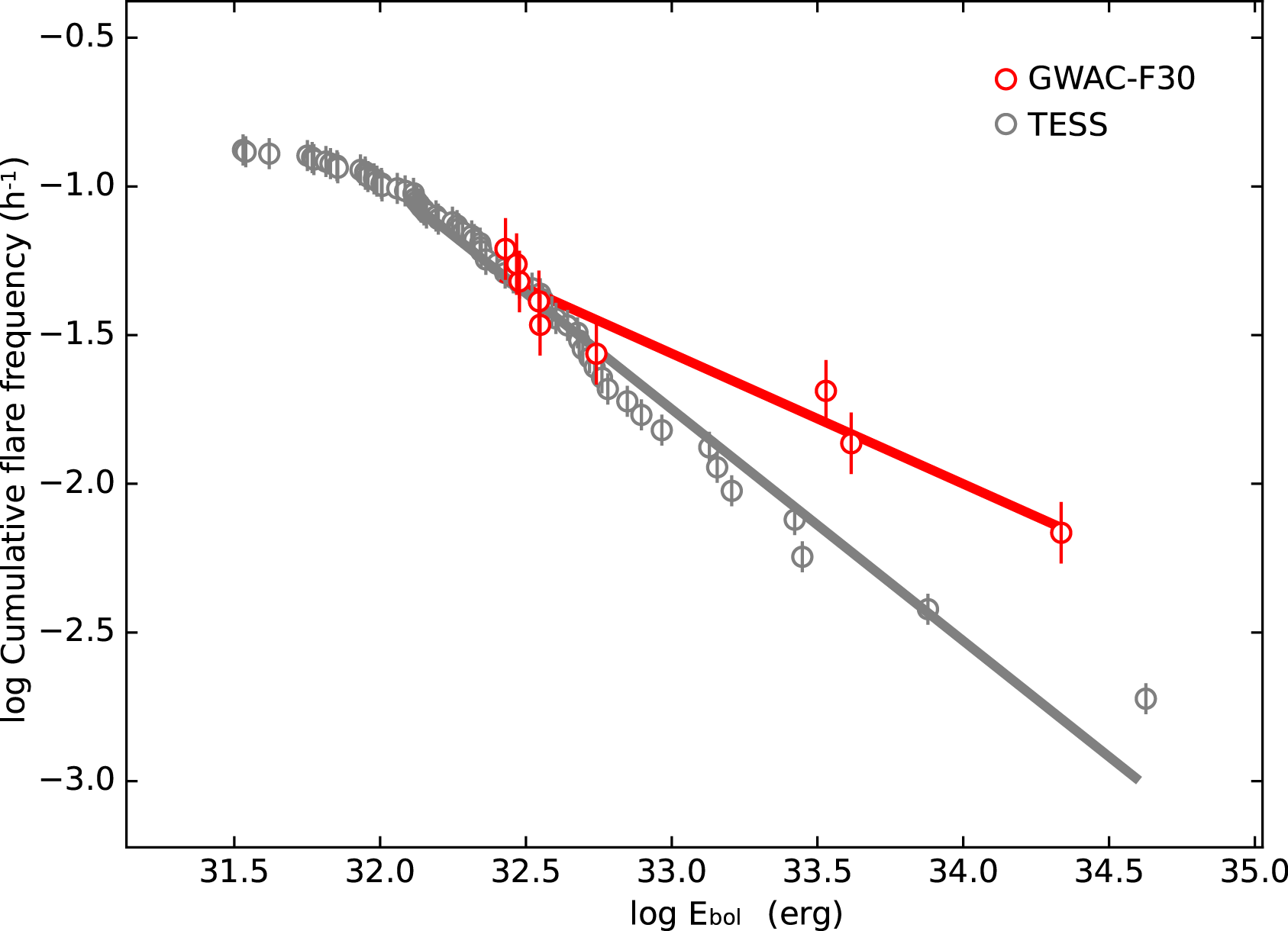}
	\caption{Cumulative flare frequency versus flare energy of GWAC-F30 and TESS. For GWAC-F30, the red circles represent the cumulative flare frequency of each flare energy, and gray circles for TESS. The red and gray lines trace the least-squares power-law fit. The vertical error bars illustrate standard errors.\label{fig:f2}}
\end{figure}

\begin{figure}[h]
	\centering
	\includegraphics[angle=0,width=110mm]{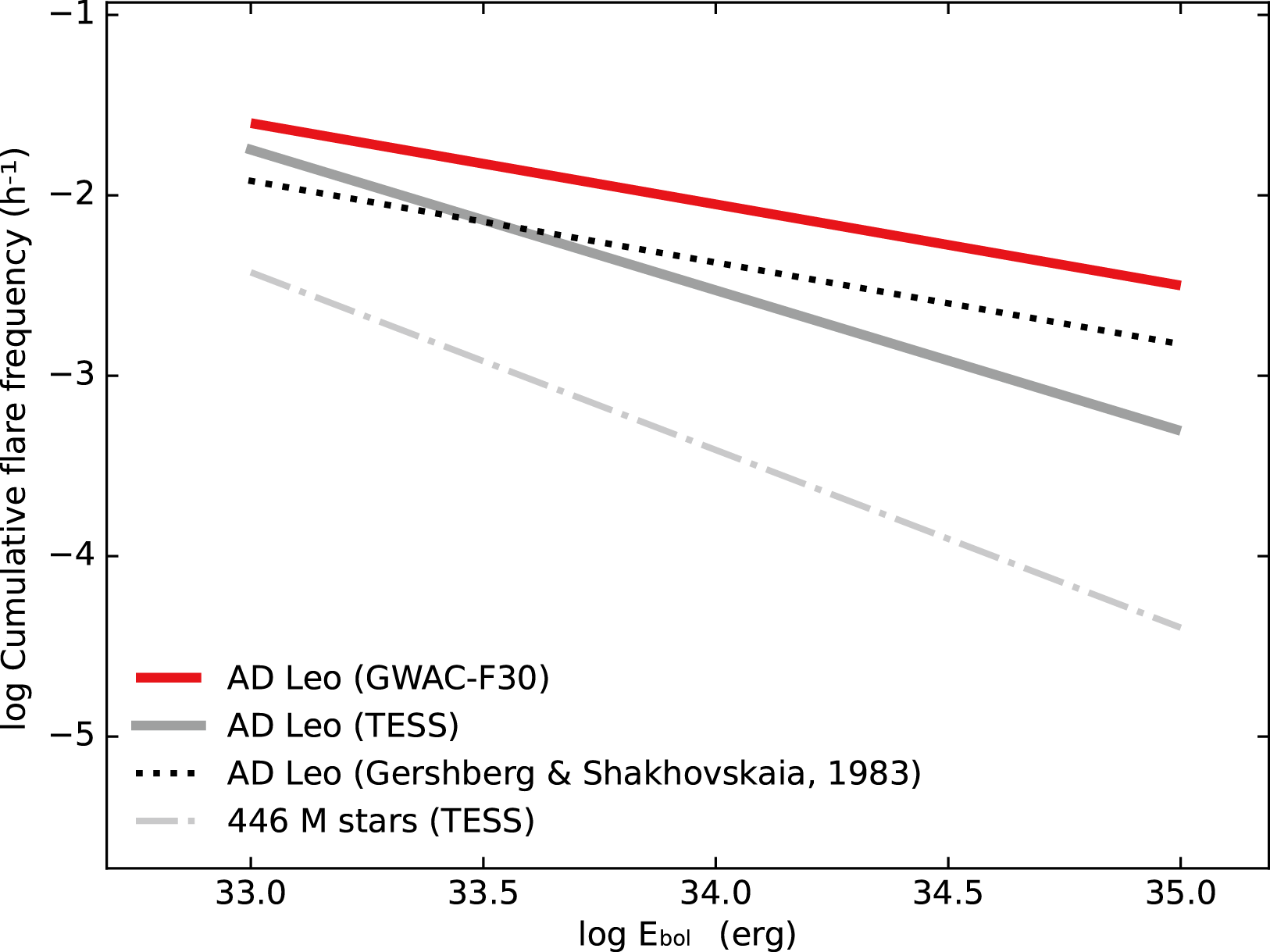}
	\caption{Comparison of AD Leo's FFDs in this work (red and gray lines) with that of the same star obtained in \cite{1983Ap&SS..95..235G} (dotted line) and the average FFD of 446 M stars getted from \cite{2020AJ....159...60G} (dash-dot line).} \label{fig:f4}
\end{figure}

Cumulative flare frequency distribution (FFD) presents the relationship of energies and burst rates of flares \citep{1972Ap&SS..19...75G,1976ApJS...30...85L,2012PASP..124..545H,2018ApJ...858...55P}. To determine the bolometric flare energies, we assume flare temperature is about 12000 K \citep{2022MNRAS.516..540R}. The fraction of the flux falls within the B band is calculated as about 0.15, and for TESS band is about 0.14 which is same with that applied in previous research \citep{2019A&A...628A..79S,2022MNRAS.516..540R}. The bolometric flare energies are calculated as B-band or TESS-band flare energies divided by the fractions in corresponding band. Both the flare energy in B or TESS band are obtained by the equivalent duration \citep{1972Ap&SS..19...75G} of the flare multiplied by the quiescent luminosity of AD Leo following the method described in \cite{2014ApJ...797..121H}. The equivalent duration is defined as the time it takes for a star to release the same amount of energy at quiescent stage as it releases during a flare. It is computed as the time integral of F$_{f}$(t)/F$_{0}$, where F$_{f}$(t) is the flux of the flare and F$_{0}$ is the flux of the star in the quiescent state \citep[see][]{1972Ap&SS..19...75G,2012PASP..124..545H,2014ApJ...797..121H}. The B-band flux of AD Leo is obtained by convolving the transmission of B filter with the quiescent spectrum of the star taken from \cite{2004A&A...414..699C}. The quiescent luminosity in B band is obtained with 4$\pi$d$^2$ multiplied by the B-band flux of the quiescent state of AD Leo, where d is 4.9 pc \citep{1991PASP..103..439G}. AD Leo’s quiescent luminosity in B band is calculated as 3.73 $\times$ 10$^{30}$ erg s$^{-1}$, and then the quiescent luminosity in TESS band is obtained as the quiescent luminosity of B band divided by 0.15 and then multiplied by 0.14. The errors of the flare energies are computed with the photometry errors. The flare amplitudes, times of rise and decay, and the total duration are also calculated (see Table 2). 

To determin AD Leo's rotation period, the TESS light curve is analyzed by the PDM function in IRAF, which is based on the Phase Dispersion Minimization method \citep{1978ApJ...224..953S}. We also analyse the decay phase of the largest flare in TESS data to determin the period of quasi-periodic pulsations(QPP; \cite{2021SSRv..217...66Z}). The top panel of Figure 8 shows the light curve of the strongest flare and the best fitting line to the decay stage. The decay phase is fitted by the function f(x)=A1*exp$^{(-x/T1)}$+A2*exp$^{(-x/T2)}$+A3, in which x and f(x) are the time and flux for each data point in light curve, respectively, and the other terms in the function are fitting parameters. And then, the fitting line is subtracted from the decay phase, obtaining the resiudual light curve (middle panel of Figure 8). In order to get the QPP period, the Lomb-Scargle method \citep{2009A&A...496..577Z} is applied to process the residual light curve. The power-period diagram is shown in bottom panel of Figure 8.

\begin{figure}[h]
	\centering
	\includegraphics[angle=0,width=110mm]{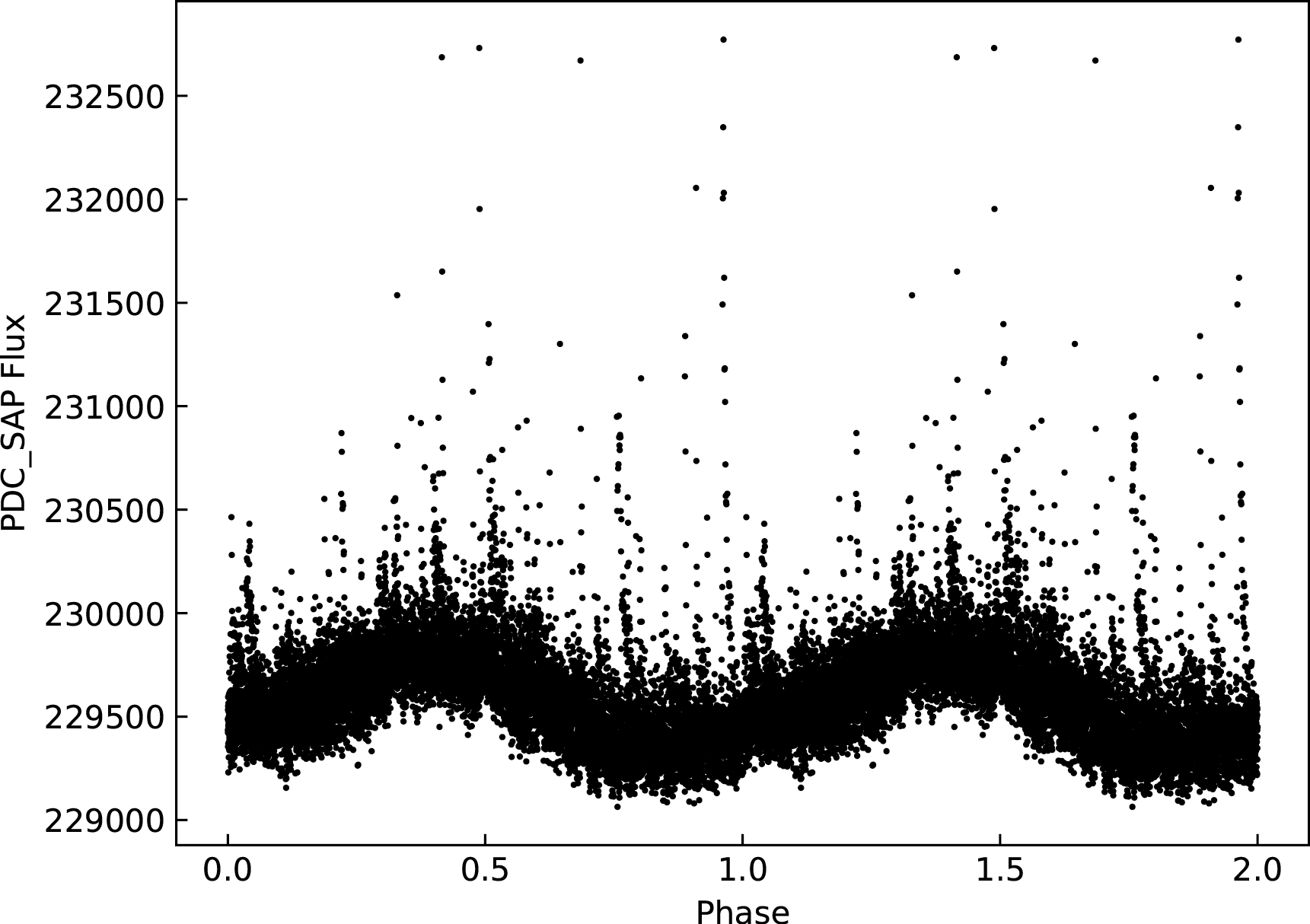}
	\caption{Phase-folded light curve from TESS with the rotation period of 2.21 day for AD Leo.} \label{fig:f4}
\end{figure}

\begin{figure}[h]
	\centering
	\includegraphics[angle=0,width=110mm]{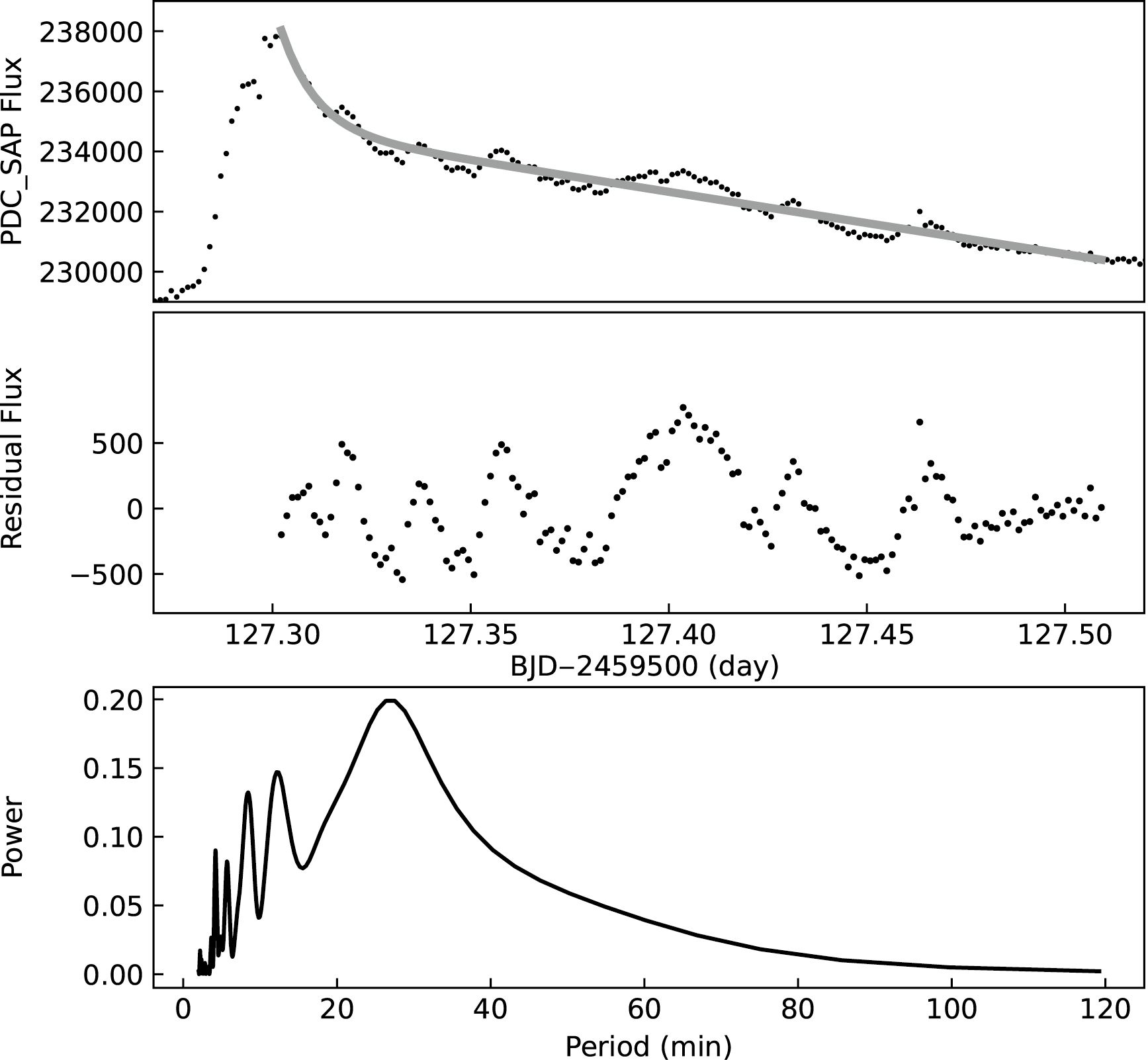}
	\caption{Top: The flare light curve with QPPs from TESS (black dot). The gray line dnotes the best fitting of the flare decay phase. Middle: Residual light curve of the flare decay stage subtracted by the fitting line. Bottom: Power-period diagram calculated with the Lomb–Scargle method.} \label{fig:f4}
\end{figure}

\section{Results and discussions}\label{sec:Results}

During the photometry observations of 146 hr to AD Leo by GWAC-F30 and 528 hr by TESS, a total of 9 and 70 flares are identified, respectively, with the method given in Section 3. Figure 3 gives light curves of the flares from GWAC-F30, and the flares in panel (f) and (g) are also identified in TESS data. The panel (a) of Figure 3 shows the light curve of the strongest flare detected by GWAC-F30 which obviously involves two different processes in the rise phase, slow and rapid rise. The two different processes may be dominated by emission-line and continuum radiations, respectively \citep{1973ApJ...185..239B}. Table 2 lists the parameters of each flare from GWAC-F30, including the flare ID, the times of the rise and decay, the total duration, the peak amplitude in magnitude and the energy. For the flares detected from TESS data, Figure 4 shows their distributions of  durations, amplitudes and energies, and the median values of the three parameters are 28.0 min, 4.3 mmag and 10$^{32.5}$ erg, respectively. The temperatures of flares in panel (f) and (g) of Figure 3 are calculated as about 38000 and 9700 K, respectively.

Figure 5 displays the AD Leo's FFD and the fitting lines of GWAC-F30 and TESS. The FFDs are fitted by the least-squares power law of log(Flare frequency) = –0.44$^{+0.15}_{-0.15}$ log(E$_{bol}$) + 12.91$^{+4.30}_{-4.30}$ and log(Flare frequency) = –0.78$^{+0.09}_{-0.09}$ log(E$_{bol}$) + 23.96$^{+2.86}_{-2.86}$, respectively. It is noted that the FFD slope of GWAC-F30 is smaller than that of TESS, which is probably due to the lower detection limit of GWAC-F30 than that of TESS. We also computed flare energies detected in the data of B-band and TESS-band in different continuum temperatures(10000, 20000 and 30000 K), to compare FFDs between B band and TESS band. The discrepancy, as shown in Figure 5, between FFDs of the two bands is becoming smaller as the temperature rises, suggesting that the discrepancy is correlated to the selected temperature.

Figure 6 shows the FFDs of AD Leo, within the same energy range,  calculated by this work and obtained from \cite{1983Ap&SS..95..235G} which identified 25 B-band flares in 1010-hr photometric observations with a 64-cm telescope, and the average FFD of 446 M-type flare stars from \cite{2020AJ....159...60G}. The FFDs of AD Leo from different observations are basically consitent, and it is apparent that the FFDs are higher than that of the average, indicating the magnetic activity on AD Leo is more active than the average activity level of these M-type stars. 

Based on TESS light curve, the rotation period is determined as 2.21$^{+0.01}_{-0.01}$ day, which is consistent with that (about 2.23 day) given in previous research \citep[e.g.,][]{2008MNRAS.390..567M,2012PASP..124..545H,2018AJ....155..192T}. Figure 7 indicates the phase diagram of AD Leo with the period of 2.21 day. Several studies have indicated that the M-type flare stars with shorter rotation periods present higher flare activity \citep[e.g.,][]{2019ApJ...873...97L,2019ApJS..241...29Y}. The rapid rotation may be the cause of the active flare activity on AD Leo. Together with that AD Leo is seen nearly pole-on, the FFD result supports that this star may has more opportunities to observe CMEs \citep{2008MNRAS.390..567M,2020A&A...637A..13M}.  In addition to the rotation cycle, \cite{2014ApJ...781L...9B} pointed out that AD Leo may have two longer chromospheric activity cycles by analyzing spectral and photometric data, approximately seven and two years, which are explained by two dynamo mechanisms acting near the surface and deep chroposphere, respectively. 

Using the Lomb–Scargle method, we analyzed the decay phase of the strongest flare on AD Leo in TESS data. QPPs with multiple cycles are well seen in the residual light curve as shown by the middle panel of Figure 8. The bottom panel in Figure 8 presents the power-period diagram, and an oscillation period of 26.5 min is determined for the QPPs. The phenomenon of QPPs have been detected both in solar and stellar flares, at least fifteen models are developed to explain the phenomenon in solar flares\citep[see review ][]{2021SSRv..217...66Z}, but which one is closest to reality is still unknown.  Up to now, multiple researches have found QPPs during flares of AD Leo and the QPP periods range from seconds to several minutes \citep{1990ApJ...353..265B,1993A&A...274..245H,2013ARep...57..603L}. For AD Leo, the QPPs detected in our study has the longest period. By analysing the short-cadence data of Kepler mission, \cite{2016MNRAS.459.3659P} finds that QPPs period are uncorrelated with stellar parameters and flare energies, indicating that the QPPs are independent of global properties of stars.

\section{Summary}\label{sec:Summary}

AD Leo was observed by GWAC-F30 in B bandpass from December, 2021 to February, 2022, obtaining 146-hr photometric data. The photometric data in sector 48 from TESS were also analyzed. A total of 9 and 70 flares are detected in GWAC-F30 and TESS data, respectively. Flare duration, amplitude and energies are calculated, and the distribution of the three properties and the FFDs are given. The FFD of AD Leo is obtained from previous research in which 1010-hr photometric data in B band is analyzed. Within the same flare energy range, the FFDs of AD Leo obtained by GWAC-30, TESS and the previous study are basically consistent. The average FFD of 446 M-type stars is obtained from previous study in which the author identified and analyzed flares based on TESS data. AD Leo's FFD is significantly higher than the average FFD of M-type, which suggests that the star's magnetic activity is more active than that of the average level of these M-type stars. We determined AD Leo's rotation period as 2.21$^{+0.01}_{-0.01}$ day by analyzing TESS light curve, and the period is consistent with that given in previous research. The rapid rotation may be the cause for AD Leo's active flare activity. During the decay phase of the most energetic flare of TESS, we identified the QPPs and determined the oscillation period of 26.5 min which is currently the longest one for AD Leo.

\begin{acknowledgements}
	This research is supported by the National Natural Science Foundation of China (Grant No. 12133003, U1831207, U1938201 and 11973055) and the Strategic Pionner Program on Space Science, Chinese Academy of Sciences (Grant No. XDA15052600).
	This paper includes data collected by the TESS mission, which are publicly available from the Mikulski Archive for Space Telescopes (MAST). Funding for the TESS mission is provided by NASA’s Science Mission directorate.
\end{acknowledgements}

%


\bibliographystyle{aasjournal}
\bibliography{papers}


\end{document}